# ChatGPT impacts in programming education: A recent literature overview that debates ChatGPT responses

Christos-Nikolaos Anagnostopoulos

University of Aegean, Cultural Technology and Communication Dpt,
Intelligent Systems lab, Lesvos, Mytilene 81100, Greece.
Email: canag@aegean.gr

**Abstract.**

This paper aims at a brief overview of the main impact of ChatGPT in the scientific field of programming and learning/education in computer science. It lists, covers and documents from the literature the major issues that have been identified for this topic, such as applications, advantages and limitations, ethical issues raised. Answers to the above questions were solicited from ChatGPT itself, the responses were collected, and then the recent literature was surveyed to determine whether or not the responses are supported. The paper ends with a short discussion on what is expected to happen in the near future. A future that can be extremely promising if humanity manages to have AI as a proper ally and partner, with distinct roles and specific rules of cooperation and interaction.

**Keywords:** ChatGPT, Programming, Computer Science, Overview, Survey Education, Ethics.

## 1. Introduction

Recently, in the field of human-computer communication and natural language processing, AI-based language models such as several versions of ChatGPT (3 and 4) have shown excellent performance in translation, question answering, inference estimation, text evaluation and summarization, but also in the field of automatic code generation and programming functions. These models have the capability to generate code and assist programmers in completing tasks more efficiently. However, there are challenges associated with AI-generated code, such as syntax errors that can hinder comprehension and functionality. Despite these challenges, the integration of AI in computer science and code programming has the potential to enhance productivity, accessibility, and problem-solving capabilities.

This paper attempts to provide a brief and recent overview that supports or negates the points that ChatGPT itself identifies as important impacts, advantages and disadvantages. Moreover, the paper concentrates some structural categories for the major impacts of them in computer science (programming and education) as identified by ChatGPT and as discussed in over 40 recent research papers (most of them published in 2022 and 2023).

The major goal is to provide a brief source of reference for researchers involved in the above scientific fields. This would support the systematic performance evaluation in the scientific community worldwide and allow developers to get familiarized with technologies that are still immature for the time being, but there are here to stay. It's now time, not to be afraid of them, but adopt with them and use them as assistive tools.

## 2. Impacts of AI generated content and ChatGPT in programming

The following subsections discuss what are the main implications of AI Generated content and ChatGPT in the field of computer science programming and programmers training. It is chosen to evaluate the main applications, advantages, disadvantages, future implications, and ethical considerations providing recent documentation and sources that support those outcomes. Thus, the following five questions were addressed to ChatGPT ver3.5, in order to identify the opinion of ChatGPT in these issues:

*Question 1: What are the most important applications of ChatGPT in programming and training of programmers?*

*Question 2: What are the major advantages of ChatGPT in programming and training of programmers?*

*Question 3: What are the most important limitations of ChatGPT in programming and training of programmers?*

*Question 4: What are the future prospects of ChatGPT in programming and training of programmers?*

*Question 5: What are ethical considerations of ChatGPT in programming and training of programmers?*

In the following five sections the responses of ChatGPT in every one of the above five questions are given in bullets in the respective tables. Each table demonstrates the recent papers identified in the literature that have a positive, negative or a partially positive/negative view. The tables are then followed by a brief discussion.

### 3.1. Question 1: Applications

Table I. The question about "most important applications" in ChatGPT and the responses are highlighted in grey.

| Question: |
|---|
| **What are the most important applications of ChatGPT in programming and training of programmers?** |

| ***ChatGPT3.5 response In bullets below:*** | *Positive view* | *Negative view* | *Partially positive and negative* |
|---|---|---|---|
| a. Assistance in code management and production | (Sallam, 2023), (Devlin, 2018), (Liu et al., 2019), (Li, 2023 | (Prather, 2023) | |
| b. Programming tutoring | (Zambrano, 2023) | (FERROUHI, 2023) (Hopkins et al., 2023) | (Iskender, 2023) (Uddin et al., 2023) |
| c. Natural Language Processing | (Serdaliyev & Zhunissov, 2023), (Gilson et al., 2023), (Temsah et al., 2023), (Stokel-Walker & Noorden, 2023), (Buriak et al., 2023), (Banerjee et al., 2023), (Ali et al., 2023) | (Qasem, 2023) | (Qasem, 2023), (Stokel-Walker & Noorden, 2023), (Sallam,2023) |
| d. Technical Documentation | (Liu et al., 2019) | | |

**a. Assistance in code management and production:** Automatically generated AI content helps developers by showing suggestions for code or code snippets, syntax checks and debugging help. Indeed, ChatGPT is an effective tool for generating computer language code snippets (Sallam, 2023; Devlin, 2018), but also for conducting comprehensive literature reviews and queries from the literature, thus helping scientists even in creating content for writing research proposal suggestions (Liu et al., 2019; Li, 2023). But in any case, if one wants to use an AI tool specifically tailored for programming, then GitHub Copilot is the optimal assistant that works with AI principles (Prather, 2023).

**b. Programming tutoring:** Content generated using AI is an excellent aid for users who are first learning computer programming languages, as it is possible to develop explanatory dialogues for various concepts and answer basic questions. ChatGPT offers process assistance for writing code, improving training in programming languages and also supporting users in programming tasks. ChatGPT has even been compared to other tools for automated programming code generation and has been found to have unique capabilities (Zambrano, 2023). In the context of research and education, ChatGPT has been found to be a tool that assists researchers in writing research papers, as mentioned above. However, it is particularly critical to keep in mind that ChatGPT several times requires the user to validate and verify the accuracy of the information it provides. The quality of ChatGPT's responses, particularly in literature reviews, literature reports and data sources (FERROUHI, 2023; Hopkins et al., 2023) has been debated and specific doubts have been raised about its accuracy.

In the field of computer programming education, ChatGPT has also been used to assist teachers in related tasks, such as grading and assessment. However, just as is the case for learners/students, there are serious questions about the risk of reducing instructors' critical thinking if they become overly dependent on ChatGPT (Iskender, 2023; Uddin et al, 2023). It is no coincidence then that the overall conclusion is that ChatGPT can damage the critical thinking ability of its users (both students and instructors) and need for continuous self-improvement if the latter do not understand the correct degree of its use and value it as the main tool of the educational process.

**c. Natural Language Processing (NLP):** ChatGPT finds particularly important applications in the field of natural language processing (NLP), which include computer science education as mentioned above, but also its assessment, automatic question answering systems, automated text generation, as well as translation between texts in different languages and types (Serdaliyev & Zhunissov, 2023).

In the broader field of NLP, ChatGPT found immediate applications in chatbots and virtual assistants (Serdaliyev & Zhunissov, 2023; Gilson et al., 2023), which is not surprising as it is itself a form of chatbot. Moreover, ChatGPT is proposed for automated paper writing applications, and text generation (text from scratch or revised versions) for research papers and articles (Gilson et al., 2023; Temsah et al., 2023).

Beyond writing papers, ChatGPT is reported to be used for research paper editing, academic instruction, and knowledge-based assessments (Stokel-Walker & Noorden, 2023; Buriak et al., 2023; Banerjee et al., 2023; Ali et al., 2023) as well as a tool for writing research proposals for funding and for academic theses (Qasem, 2023). But, as will be discussed in the section of ethical issues raised, a great concern is identified about an increase in plagiarism and over-reliance on ChatGPT (Qasem, 2023) leading to limitations in analytical capabilities (Stokel-Walker & Noorden, 2023) and for this case, the necessity of human verification before submitting proposals or a thesis is also emphasized (Sallam, 2023).

**d. Technical Documentation:** Artificial intelligence text generation models are capable of assisting in the completion of technical documentation and explanatory text for projects or software applications. Demonstrated success in this area brings to the fore the importance of design choices and raises questions about the source of recent improvements in the performance of language models (Liu et al., 2019). The attention mechanism, as presented in (Vaswani et al., 2017), is the one that has played the most crucial role in the development of these language models. Therefore, by incorporating the power of AI language models, software projects and applications can benefit from the automated production of technical documentation that should always accompany a software project thus increasing its sustainability and updating it with fully documented new releases.

### 3.2. Question 2: Advantages

Table II. The question about "major advantages" of ChatGPT and the responses are highlighted in grey.

| Question: What are the major advantages of ChatGPT in programming and training of programmers? | | | |
|---|---|---|---|
| *ChatGPT3.5 response In bullets below:* | *Positive view* | *Negative view* | *Partially positive and negative* |
| a. Increase of the software developer productivity and rapid idea prototyping | (Sallam, 2023), (Haque & Li, 2023), (Qadir, 2022), (Kreitmeir & Raschky, 2023), (Haque & Li, 2023) | | |
| b. Continuous Learning | (Nascimento, 2023), (Lund & Wang, 2023), (Irons, 2023) | | |

**a. Increase of the software developer productivity and rapid idea prototyping:** AI Generated Content can speed up development tasks, reduce coding errors, and enhance the overall productivity of software engineers. It also enables individuals with little or no programming experience to engage with computer science concepts and coding. In particular, ChatGPT can be used to speed up software development tasks and enhance the overall productivity of engineers engaged in these tasks (Sallam, 2023; Haque & Li, 2023). As mentioned in the previous section, ChatGPT helps in computer code generation or software prototypes, and with its proper utilization, software engineers can save time from low-level tasks that require significant effort and focus on higher-level tasks and experimental design (Sallam, 2023), get explanations, feedback and acquire realistic virtual simulations that may contribute to hands-on learning experiences (Qadir, 2022). A very good example is that by using ChatGPT, developers can take advantage of its ability to easily produce code at a fairly satisfactory level, fill in any missing parts and/or update it, or even develop it in different programming languages (Kreitmeir & Raschky, 2023). Moreover, an important capability of ChatGPT is the provision of explanations and suggestions that can help reduce coding errors and optimize the quality of the code being programmed (Haque & Li, 2023).

**b. Continuous Learning:** ChatGPT can also assist in the formation of collaborative workflows during the implementation of an IT project or during the production of software, forming an enhanced human-machine interaction (Nascimento, 2023). At the same time, ChatGPT's capabilities in natural language processing, beyond the creation of new content, have been reported to offer significant potential in information retrieval, cataloging and indexing, and metadata creation (Lund & Wang, 2023). In addition to the above, it is important to add its flexibility and interactivity that has been found to increase the productivity of scientists and their ability to access new knowledge and skills (Irons, 2023).

### 3.3. Question 3: Limitations

Table III. The question about "most important limitations" of ChatGPT and the responses are highlighted in grey.

| Question: What are the most important limitations of ChatGPT in programming and training of programmers? | | | |
|---|---|---|---|
| **ChatGPT3.5 response** *In bullets below:* | *Positive view* | *Negative view* | *Partially positive and negative* |
| a. Lack of Context | (Wang et al., 2021), (Khoshafah, (2023), (Aljanabi et al., 2023), (Buriak et al., 2023) | | |
| b. Security concerns | (Flanagin, 2023) (Buriak et al., 2023) | | |
| | | | |
| c. Overconfidence or excessive trust in AI | (Tang, 2023) | | |

**a. Lack of Context:** This is the most important limitation of AI Generated Content, since AI still cannot fully understand the context of specific codebases or assets, leading to potential inaccuracies and inappropriate promptings.

Wang et al. (2021) discuss the limitations of current methods in processing code snippets, neglecting the special characteristics of programming languages and token types, which can result in suboptimal understanding and generation tasks. Khoshafah (2023) denotes issues related to the translation accuracy of ChatGPT and highlights limitations in understanding domain-specific terminology and cultural context, which can affect the accuracy of translations. Aljanabi et al. (2023) mentions that ChatGPT may not fully understand the nuances of security vulnerabilities, reverse engineering, or malware analysis, potentially leading to inaccurate or less useful information in coding. Additionally, it mentions that ChatGPT may not be able to handle certain types of queries, such as mathematical calculations.

**b. Security Concerns:** AI models like ChatGPT might inadvertently generate code with security vulnerabilities, which could be exploited by malicious actors. Security concerns arise in the use of ChatGPT in computer science applications. The potential for misuse, such as using ChatGPT to cheat on assignments, write essays, or take exams, including computer science and programming exams, has been identified. The inclusion of ChatGPT as a bylined author in scientific articles has raised concerns about the integrity of scientific publication and the indexing of nonhuman "authors". The lack of accountability and responsibility associated with AI tools like ChatGPT has prompted the development of policies by journals and organizations to address these issues (Flanagin, 2023). Additionally, the limitations of ChatGPT, such as its lack of context and inability to understand new information or generate deep analysis, restrict its utility in writing up-to-date reviews, perspectives, and introductions (Buriak et al., 2023). It is crucial to address these security concerns and ensure responsible use of ChatGPT in computer science to maintain the integrity of scientific research and programming education.

**c. Overconfidence or excessive trust in AI:** Dependence on AI-generated code without proper understanding certainly would negatively affect developers' growth. While AI technologies like low-code development platforms offer efficiency and productivity benefits, relying solely on AI-generated code may limit developers' understanding of underlying concepts and principles. This lack of understanding can hinder their ability to troubleshoot, debug, and optimize code, resulting in less robust and maintainable solutions. Additionally, the lack of comprehension may impede developers' growth and hinder their ability to adapt to new technologies and challenges (Tang, 2023). Therefore, it is important to strike a balance between implementing AI technologies and ensuring developers' comprehensive understanding of code and underlying principles.

### 3.4. Question 4: Future Prospects

Table IV. The question about "future prospects" of ChatGPT and the responses are highlighted in grey.

| Question: What are the future prospects of ChatGPT in programming and training of programmers? | | | |
|---|---|---|---|
| *ChatGPT3.5 response In bullets below:* | *Positive view* | *Negative view* | *Partially positive and negative* |
| a. AI languages customization | Not found specifically for ChatGPT but for AI Generated Content in general: (Devlin,2018), (Liu et al., 2019), (Brown et al., 2020) | | |
| b. Collaborative operations | Not found specifically for ChatGPT but for AI Generated Content in general: (Mikhaylov et al., 2018) | | |
| c. | Not found specifically for ChatGPT but for AI Generated Content in general: (Holtzman et al., 2019), (Ziegler, 2019), (Brown et al., 2020) | | |

**a. AI languages customization:** Future iterations of AI language models would certainly be tailored to specific programming languages and frameworks, providing more specialized support. As an example, Devlin (2018) introduced BERT (Bidirectional Encoder Representations from Transformers), a language representation model that can be fine-tuned for various tasks. BERT was designed to pretrain deep bidirectional representations from unlabeled text, and it has been shown to achieve state-of-the-art results on multiple natural language processing tasks (Devlin, 2018). Additionally, Liu et al. (2019) presented RoBERTa, a robustly optimized BERT pretraining approach. While the latter does not directly address the customization of AI language models for specific programming languages and frameworks, it provides insights into the advancements and optimizations in BERT-based models. More recently, Brown et al. (2020) discusses the scaling up of language models and its impact on task-agnostic, few-shot performance giving evidence about the performance improvements achieved by scaling up language models.

**b. Collaborative operations:** AI models may facilitate collaborative programming, enabling developers to work together more effectively. Back in 2018, Mikhaylov et al. (2018) discussed the challenges of cross-sector collaboration in the public sector when adopting AI and data science and programming. The authors described challenges such as information silos, lack of resources, and collaborative culture. In addition, the authors highlighted the divergent approaches to managing risk in the public and private sectors. Since the political and market risks may not easily align, challenges may be created in collaborative programming efforts using AI models. Moreover, they continue mentioning that cross-sector collaborations, although popular, entail serious management challenges that hinder their success and may impact the effectiveness of collaborative programming facilitated by AI models.

**c. Self-Correcting Models:** AI models would eventually become better at recognizing and fixing their mistakes, leading to more accurate responses. AI language models have shown remarkable progress in various NLP tasks, but their ability to recognize and fix mistakes is an ongoing challenge. However, with the exploration of decoding strategies, fine-tuning approaches, and reward learning, there is potential for self-correcting models in ChatGPT. By addressing the limitations and leveraging advancements in the field, AI models could improve their error recognition and correction capabilities, ultimately leading to more accurate responses. Specifically, to address the challenge of mistake recognition, researchers have explored different decoding strategies for text generation from language models. (Holtzman et al., 2019) propose Nucleus Sampling, a method that draws higher quality text from neural language models compared to traditional decoding strategies. By truncating the unreliable tail of the probability distribution and sampling from the dynamic nucleus of tokens, Nucleus Sampling aims to avoid text degeneration and generate more coherent and diverse responses.

In addition, recent studies have examined the fine-tuning of language models by exploiting user preferences and reinforcement learning models to improve the accuracy and quality of responses. For instance, Ziegler (2019) proposes a reward-based learning in natural language dialogues, ultimately creating a model of retributive learning implemented by asking users questions. This approach is a successful application of reinforcement learning in real-world dialogue scenarios where rewards are determined by human judgment through the incorporation of human preferences and feedback that could be incorporated in ChatGPT future versions.

In support of the above article, Brown et al. (2020) very emphatically cited the weaknesses that NLP systems faced in performing new linguistic tasks when given few examples or simple instructions. In their article they argue that AI models are not yet at the same level of error recognition and correction capabilities as a human user, but they acknowledge that this will change radically in the near future with improved decoding strategies, such as kernel sampling, and the incorporation of sophisticated adaptation and reward learning.

### 3.5. Question 5: Ethical Considerations

Table V. The question about "ethical considerations" of ChatGPT and the responses are highlighted in grey.

| Question: What are the ethical considerations of ChatGPT in programming and training of programmers? | | | |
|---|---|---|---|
| *ChatGPT3.5 response In bullets below:* | *Positive view* | *Negative view* | *Partially positive and negative* |
| a. Plagiarism, proper citation and acknowledgement | (Li,2022), (Gibea and Uszkai, 2023), (Svetlova, 2022), (Minkkinen et al., 2022), (Almarzoqi & Albakjaji, 2022), (Lu et al., 2023), (Haonan et al., 2023) | | |
| b. Bias and Fairness | Not found specifically for ChatGPT but for AI Generated Content in general: (Taddeo & Floridi, 2018), (Jobin & Vayena, 2019), (Caliskan et al., 2017), (Dixon et al., 2018), (Garg et al., 2018), (Amodei et al., 2016) | | |

**a. Plagiarism, proper citation and acknowledgement:** This is one of the most important ethical issues that arises as the code generated by AI should be recognized as automatically generated and programmers should not appropriate these results without critical processing (Taddeo and Floridi, 2018).

To this end, Li (2022) addresses the issue of source code intellectual property and the necessity for accurate attribution of code to its owner/developer, issues that are crucial in the software and application implementation cycle and in the broader fields of software forensics and software quality analysis. In the same article, the author acknowledges that the current methods of rendering the authoring of a code can be copied by other users, which means that the same phenomena can occur in ChatGPT. Furthermore, Gibea and Uszkai (2023) summarize in their article the growing concern of the research community about the development and deployment of AI systems and point out the need for a common code of ethics to be established by international organizations in response to these concerns. In addition, Svetlova (2022) argues that AI ethics should also take into account all systemic implications arising from the use of AI, especially the systemic effects on developers' codes of conduct. All of the above argues that the appropriate performance of an automatically generated code is critical to address potential systemic risks, but also to ensure the appropriate development of AI itself.

Similarly, the article by Jobin and Vayena, (2019), highlights the new landscape of ethical guidelines for AI with both meaningful analysis and appropriate implementation strategies. The importance of adhering to ethical considerations, among which is the appropriate attribution of the rights of a code segment, is reemphasized and also for the development of AI systems. In the same vein, Taddeo and Floridi (2018) stress the importance of having an ethical framework for exploiting the potential of AI, but where control remains strictly with the human agent. Furthermore, Minkkinen et al. (2022) present for the first time the use of AI in process evaluations in administrative, social and governmental issues, proposing standardized metrics for the responsible use of AI, including the proper performance of AI-generated code.

The important question "Does AI own intellectual property rights in content produced by it?" is answered in the article by Sharma & Sony (2020). The article highlights the uncertainty that arises on this issue, but again clearly supports the need for appropriate attribution. In addition, Díaz-Noci (2020) went a little further into the question and presented the legal implications of news produced using AI systems and what happens to intellectual property rights. Although the article is mainly concerned with journalistic practices, it highlights the challenges in determining the rights of the author, and again strongly supports the need for appropriate attribution of AI-generated content.

Very interesting is also the article by Almarzoqi and Albakjaji (2022), who even explored the possibility of patenting products resulting from AI, while they also examined the challenges posed by intellectual property laws in this context. As in the previous article, the need for appropriate attribution to the creator (AI) is emphasized. Wowever, it is recognized that the legal frameworks that describe and regulate intellectual property in innovations/new inventions are not ready to patent something that is purely generated by AI.

More recently, Lu et al. (2023) propose the idea of collecting patterns to assist in the design of responsible AI systems. Their article highlights the need to address the challenges of responsible AI operation, part of which includes the appropriate rendering of the code generated by AI. Similarly, Haonan et al. (2023) discuss copyright protection and accountability of creative AI. Although they focus on intellectual property rights, their work highlights the need for attribution and accountability in the context of AI-generated works.

**b. Bias and Fairness:** Another important ethical aspect is the presence of bias in the training data. In programming, this would lead to biased or discriminatory code suggestions, and it is important to focus on efforts to mitigate such issues.

It should be emphasized that the impact of bias in training data that leads to discriminatory code suggestions from AI generated text or code, was early underlined from the scientific community. For instance, (Caliskan et al., 2017) discuss how text corpora contain recoverable and accurate imprints of historic biases, including biases towards race or gender. This suggests that biases present in the training data can be reflected in the output of AI models, potentially leading to biased or discriminatory code suggestions. Aligned with the above assumptions, Dixon et al. (2018) demonstrated how imbalances in training data can lead to unintended bias in resulting models. This implies that if the training data contains biased or discriminatory patterns, the AI model's code suggestions may also exhibit such biases.

Moreover, Garg et al. (2018) highlighted the presence of gender and ethnic stereotypes in word embeddings trained on text data. Word embeddings, which are widely used in natural language processing tasks, can amplify biases present in the training data. This suggests that biases in training data can affect the behavior of AI models, including code suggestions.

To sum up, ChatGPT has shown potential as a code assistance tool, aiding in generating computer codes and supporting researchers in coding tasks developers should be aware when they are interacting with AI systems and understand the model's limitations to make informed decisions and take responsibility for thoroughly reviewing and testing the output. To this end, Amodei et al. (2016) focuses on AI safety and the potential risks associated with forward-looking applications of AI. While it does not directly address the need for developers to understand AI model limitations, it highlights the importance of considering safety in the design and deployment of AI systems.

## 3. Discussion – outlook for the future

The next versions of ChatGPT, which will certainly be more comprehensive instances of AI-generated content tools, will significantly improve the automatic chat capabilities. As an AI-based language model, ChatGPT is designed to be trained by its interactions with humans and appropriately adapt its responses. As such, it is reasonable to expect that future versions of ChatGPT, will be more comprehensive and reliable, as they will have gathered a greater amount of data and experience. This accumulation of knowledge will allow it to provide even more accurate and appropriate answers for the subjects for which questions are asked, both in general discussions and specifically in areas such as computer science education and code generation.

On a technical level, the ChatGPT developers will certainly incorporate new updates that will include improvements in natural language processing, increased ability to understand user queries and the ability to generate responses that are more coherent and closer to everyday human communication in natural language. In addition, it is reasonable to assume that the developers have addressed any limitations or biases that were identified in previous versions, ensuring an ever-increasing impartiality of ChatGPT.

In addition, the next era for ChatGPT could include the integration of new features or functions. As a representative example, a feature has already been launched where a feedback mechanism is incorporated that allows users to provide feedback on the quality of responses. This feedback loop continuously improves ChatGPT based on user feedback, resulting in a more personalized and satisfying user experience. It is also expected that ChatGPT will be integrated with different AI models through artificial agents. Such collaborations will lead to a more diverse range of answers in computer science to create code, increase the knowledge base and improve problem-solving capabilities (Neil et al., 2022) and (Beiqi et al., 2023).

However, the human factor must play an important role in the final decisions and judgements. In particular, in the field of computer science, programmers should not rely solely on the answers provided by an automated system but should instead combine them with their own expertise and experience and should continuously cultivate their knowledge, improve their skills and remain actively involved in the continuous and lifelong learning of (Christen et al., 2023).

### Data availability statement